\documentclass[11pt]{article}
\usepackage{graphicx,amsmath,bm, amsthm,mathrsfs,amssymb, braket, verbatim}
\usepackage[usenames]{color}
\usepackage{ulem,mathtools}
\usepackage{authblk}
\usepackage{cite}
\usepackage{tcolorbox}
\usepackage[section]{placeins}
\usepackage{placeins}

\newcommand{\ee}{\end{equation}}

\newcommand{\reff}[1]{(\ref{#1})}
\newcommand{\beq}{\begin{equation}}
\newcommand{\eeq}[1]{\label{#1}\end{equation}}
\newcommand{\beqa}{\begin{eqnarray}}
\newcommand{\eea}{\end{eqnarray}}
\newcommand{\eeqa}[1]{\label{#1}\end{eqnarray}}
\newcommand{\beg}{\begin{equation*}}
\newcommand{\eeg}{\end{equation*}}

\newcommand{\bsplit}{\begin{split}}
\newcommand{\esplit}{\end{split}}

\usepackage{subfig} %To put many figures side by side
\usepackage{circuitikz} %To draw the electrical circuits
\usepackage[capposition=bottom]{floatrow} %To add notes under a figure
\usepackage{epigraph} %For quotes at the beginning of sections and text
\usepackage{appendix}
\usepackage{rotating}
\allowdisplaybreaks

\title{Enlarging the symmetry of pure $R^2$ gravity, BRST invariance and its spontaneaous breaking}

\author[]{Ariel Edery\thanks{aedery@ubishops.ca}}

\affil[]{Department of Physics and Astronomy, Bishop's University, 2600 College Street, Sherbrooke, Qu\'{e}bec, Canada, J1M 1Z7.\vspace{1em}}

\begin{document}
\date{}
\maketitle
\begin{abstract}
Pure $R^2$ gravity was considered originally to possess only global scale symmetry. It was later shown to have the larger restricted Weyl symmetry where it is invariant under the Weyl transformation $g_{\mu\nu} \to \Omega^2(x)\, g_{\mu\nu}$ when the conformal factor $\Omega(x)$ obeys the harmonic condition $\Box \Omega(x)=0$. Restricted Weyl symmetry has an analog in gauge theory. Under a gauge transformation $A_{\mu}\to A_{\mu} + \frac{1}{e}\partial_{\mu} f(x)$,  the gauge-fixing term $(\partial_{\mu}A^{\mu})^2$ has a residual gauge symmetry when $\Box f=0$.  In this paper, we consider scenarios where the symmetry of pure $R^2$ gravity can be enlarged even further. In one scenario, we add a massless scalar field to the pure $R^2$ gravity action and show that the action becomes on-shell Weyl invariant when the equations of motion are obeyed. We then enlarge the symmetry to a BRST symmetry where no on-shell or restricted Weyl condition is required. The BRST transformations here are not associated with gauge transformations (such as diffeomorphisms) but with Weyl (local scale) transformations where the conformal factor consists of a product of Grassmann variables.  BRST invariance  in this context is a generalization of Weyl invariance that is valid in the presence of the Weyl-breaking $R^2$ term. In contrast to the BRST invariance of gauge theories like QCD, it is not preserved after quantization since renormalization introduces a scale (leading to the well-known Weyl (conformal) anomaly).  We show that the spontaneous breaking of the BRST symmetry yields an Einstein action; this still has a symmetry which is also anomalous. This is in accord with previous work that shows that there is conformal anomaly matching between the unbroken and broken phases when conformal symmetry is spontaneously broken.        
\end{abstract}

\setcounter{page}{1}
\newpage
\section{Introduction}\label{Intro}
Pure $R^2$ gravity ($R^2$ alone with no additional $R$ term) is unique among quadratic gravity theories as it is unitary and moreover has been shown to be conformally equivalent to Einstein gravity with non-zero cosmological constant and massless scalar field \cite{Lust1,Lust2,AE1,Lust3,AE2} (though in a Palatini formalism one can avoid having a massless scalar \cite{AE5}). It has been known for a long time that it is invariant under the global scale transformation $g_{\mu\nu}\to \lambda^2\,g_{\mu\nu}$ where $\lambda$ is a constant. It was later discovered to possess a larger symmetry than global scale symmetry called restricted Weyl symmetry \cite{AE3} where it is invariant under the transformation $g_{\mu\nu} \to \Omega^2(x) \,g_{\mu\nu}$ when the conformal factor $\Omega(x)$ obeys the harmonic condition $\Box \Omega= g^{\mu\nu}\nabla_{\mu}\nabla_{\nu} \Omega=0$. The conformal factor $\Omega(x)$ is therefore not limited to being a constant. The aforementioned equivalence between pure $R^2$ gravity and Einstein gravity with cosmological constant was then interpreted in a new light: it occurs when the restricted Weyl symmetry is spontaneously broken \cite{AE1,AE2}. In the broken sector, the Ricci scalar of the  background (vacuum) spacetime has $R \ne 0$  which excludes a flat background. This is why the equivalence requires a non-zero cosmological constant on the the Einstein side. The unbroken sector which has an $R = 0 $ vacuum (background) has no relation to Einstein gravity. In fact, it has been shown that a linearization of pure $R^2$ gravity about Minkowski spacetime  does not yield gravitons but only a propagating scalar \cite{Lust3}; simply put,  pure $R^2$ gravity does not gravitate about a flat background \cite{Lust3}. However, it was later shown that if one includes a non-minimally coupled scalar field in the restricted Weyl-invariant action and the field acquires a non-zero VEV, then the theory can gravitate about flat spacetime \cite{AE2,AE6}.  Various aspects of restricted Weyl symmetry, it spontaneous breaking as well as its role in critical gravity were then explored further in \cite{AE2,AE3, AE4,AE5, Oda1,Oda2}

Restricted Weyl symmetry has an analog in gauge theory. The gauge-fixing term $(\partial_{\mu}A^{\mu})^2$ is invariant under the gauge transformation $A_{\mu}\to A_{\mu} +\frac{1}{e} \partial_{\mu} f(x)$ only when the arbitrary smooth function $f(x)$ obeys the condition $\Box f=0$ where $\Box$ here represents the flat space d'Alembertian.   Therefore, the gauge-fixing term has a residual gauge symmetry when  $\Box f=0$  is satisfied \cite{Schwartz}. This is the analog to the restricted Weyl symmetry of pure $R^2$ gravity when the conformal factor $\Omega(x)$ satisfies $\Box \Omega=0$. As we will see, this analogy is fruitful as it provides a bridge to the BRST symmetry of pure $R^2$ gravity.  Recent work on the BRST invariance of other gravitational theories can be found in  \cite{Oda3, Prinz, Berez}. 

In this paper, we consider scenarios where the symmetry of pure $R^2$ gravity is enlarged further. We show that when a massless scalar field is added to pure $R^2$ gravity, the action becomes Weyl invariant when the equations of motion are satisfied. No separate external condition is required to be imposed on the conformal factor $\Omega(x)$ as this occurs naturally via the equations of motion. One passes from restricted Weyl invariance to on-shell Weyl invariance. One can then enlarge the symmetry further to include BRST symmetry.  In analogy with the BRST invariance in gauge theories in the presence of a gauge-fixing term, we establish BRST invariance in the presence of the Weyl-breaking pure $R^2$ gravity term. The BRST transformations here are not associated with gauge transformations (such as diffeomorphisms) but are a generalization of Weyl (local scale) transformations where the conformal factor is composed of Grassmann variables. Therefore, in contrast to the BRST invariance in gauge theory, it is anomalous since renormalization introduces a scale (leading to the well-known Weyl (conformal) anomaly). We show that the spontaneous breaking of the BRST symmetry yields an Einstein action with its own symmetry that is also anomalous. This is in agreement with previous work where it was shown  that when conformal symmetry is spontaneously broken there is conformal anomaly matching in the unbroken and and broken phases \cite{Schwimmer,Bizet}. 

The paper is organized as follows. In section 2, we obtain the on-shell Weyl invariance of pure $R^2$ gravity when a massless scalar field is included in the action. In section 3, we obtain the BRST invariance of pure $R^2$ gravity. In section 4, we show that the spontaneous breaking of the BRST symmetry yields an Einstein action and that there is a quantum anomaly in both the unbroken and broken sectors. We conclude with section 5 where we summarize our results, provide further physical insights and discuss directions for future work. We relegate to Appendix A some technical details on the symmetry of the Einstein action.  

\section{Pure $R^2$ gravity plus a massless scalar: from restricted to on-shell Weyl invariance}
The action of pure $R^2$ gravity is given by 
\beq
S = \int \sqrt{-g}\, d^4x \, \alpha\, R^2 
\eeq{RSquared}
where $R$ is the Ricci scalar and $\alpha$ a dimensionless constant. This action is restricted Weyl invariant i.e. it is invariant under the Weyl transformation $g_{\mu\nu} \to \Omega^2(x) \,g_{\mu\nu}$ if the conformal factor $\Omega(x)$ obeys the condition $\Box \Omega(x)=0$. This invariance stems from the fact that $R \to R/\Omega^2$ when $\Box\Omega(x)=0$. As already mentioned, this implies that pure $R^2$ gravity has a greater symmetry than global scale symmetry (where $\Omega(x)$ would have to be a constant). 

We now show that pure $R^2$ gravity can be Weyl-invariant on-shell when a minimally coupled real massless scalar field is added to the action. Here, the condition $\Box \Omega(x)=0$ is not imposed as an external condition but satisfied automatically by the equations of motion. The action of pure $R^2$ gravity with a minimally coupled real massless scalar field $\phi$ is given by
\beq
S_a = \int \sqrt{-g}\, d^4x \, \big(\alpha\,R^2 -\dfrac{1}{2}\,g^{\mu\nu}\,\partial_{\mu}\phi \,\partial_{\nu}\phi\big)
\eeq{Sa}
where $\phi(x)$ is a real scalar field. Under the Weyl transformation $g_{\mu\nu} \to e^{-2\,\epsilon\,\phi} \,g_{\mu\nu}$, where $\epsilon$ is a real constant, the Ricci scalar transforms as 
\beq
R\to  R \,e^{2 \epsilon \phi}- 6 \,e^{3 \epsilon\,\phi}\,\Box (e^{-\epsilon\,\phi})
\eeq{R2}
and $\sqrt{-g} \to e^{-4\,\epsilon\,\phi} \,\sqrt{-g}$ so that action \reff{Sa} transforms to
\begin{align}
S_b=\int \sqrt{-g}\, d^4x \, \Big(\alpha\,\big(R^2-12 \,R\,e^{\epsilon\,\phi}\, \Box (e^{-\epsilon\,\phi})+36\, e^{2\epsilon\,\phi}  \,(\Box (e^{-\epsilon\,\phi}))^2\big) -\dfrac{1}{2}\,e^{-2\,\epsilon\,\phi}\,g^{\mu\nu}\,\partial_{\mu}\phi \,\partial_{\nu}\phi \Big)\,.
\label{S2}
\end{align}
The equations of motion yield $\Box (e^{-\epsilon\,\phi})=0$. Therefore, when the equations of motion are satisfied, the above action reduces to
\begin{align}
S_c=\int \sqrt{-g}\, d^4x \, \Big(\alpha\,R^2 -\dfrac{1}{2}\,g^{\mu\nu}\,\partial_{\mu}{\psi}\,\partial_{\nu}\psi \Big)
\label{S3}
\end{align}
where $\psi$ is a real massless scalar field (related to the old scalar $\phi$ via $\psi=e^{-\epsilon\,\phi}/\epsilon$). Note that the equation of motion for $\psi$ is $\Box \psi=0$ which is equivalent to $\Box (e^{-\epsilon\,\phi})=0$ and consistent with what we previously obtained. We therefore recover pure $R^2$ gravity with a minimally coupled real massless scalar field $\psi$. What happened here is that the restricted Weyl condition $\Box\, \Omega=0$ with $\Omega= e^{-\epsilon\,\phi}$ did not have to be imposed as a separate condition because it was satisfied automatically by the equations of motion. In short, pure $R^2$ gravity became Weyl invariant on-shell in the presence of a massless scalar field. It passed from restricted Weyl invariance to on-shell Weyl invariance. 
         
\section{BRST invariance of pure $R^2$ gravity}

Before discussing BRST invariance in the case of pure $R^2$ gravity, let us first recall how BRST invariance works in gauge theories in Minkowski spacetime. For illustrative purposes, we will consider the case of scalar QED. The Abelian version of the Faddeev-Popov Lagrangian is then given by \cite{Schwartz}
\begin{align}
\mathcal{L}=-\dfrac{1}{4}\,F_{\mu\nu}^2 -(D^{\mu} \phi_a^{*})(D_{\mu} \phi_a) -m^2\,\phi_a^{*}\,\phi_a -\dfrac{1}{2 \,\xi}(\partial_{\mu}\,A^{\mu})^2 +\bar{c}\, \Box c
\label{QED}
\end{align}
where $c(x)$ and $\bar{c}(x)$ are independent Grassmann-valued fields, $\phi_a$ are a set of complex scalar fields and $D^{\mu}$ is the usual covariant derivative.  The gauge fixing term, $\frac{1}{2 \,\xi}(\partial_{\mu}\,A^{\mu})^2$ breaks the gauge symmetry since it is not invariant under the transformation $A_{\mu}\to A_{\mu} + \frac{1}{e} \,\partial_{\mu} f(x)$ where $f(x)$ is an arbitrary function. However, it has a residual symmetry: it is invariant if $f(x)$ obeys the condition $\Box f=0$. As previously mentioned, this residual symmetry is the analog of restricted Weyl symmetry in pure $R^2$ gravity. 

The equation of motion for $c(x)$  is $\Box c=0$. Consider the gauge transformation with $f(x)= \theta\, c(x)$ for arbitrary Grassmann number $\theta$. Then, if the equation of motion for $c$ is satisfied, the scalar QED Lagrangian \reff{QED} is invariant under the following transformations 
\begin{align}
& A_{\mu} \to A_{\mu} + \frac{1}{e}\, \theta\, \partial_{\mu} c(x) \nonumber\\
& \phi_a(x)\to e^{i \theta \,c(x)}\,\phi_a(x)=\phi_a(x) + i\theta\,c(x) \phi_a(x)\,.
\label{Gauge}
\end{align}
In other words, the equation $\Box f=\theta\, \Box c = 0$ is automatically satisfied on-shell and does not have to be imposed as a separate condition. This is similar to what we saw in the previous section for pure $R^2$ gravity which was invariant under $g_{\mu\nu} \to \Omega^2 \, g_{\mu\nu}$ with $\Omega= e^{-\epsilon \phi}$  when the equations of motion were satisfied. 

If the equation of motion for $c$ is not used, the only term in the Lagrangian \reff{QED} which is not invariant under the transformation \reff{Gauge} is $(\partial_{\mu}A^{\mu})^2$ which transforms as
\begin{align}
(\partial_{\mu}A^{\mu})^2 \to (\partial_{\mu}A^{\mu})^2  + \dfrac{2}{e} (\partial_{\mu}A^{\mu}) (\theta \Box c)
\label{Gauge2}
\end{align}
where we used the fact that $\theta^2=0$ since $\theta$ is Grassmann. Now, if under \reff{Gauge} we also have $\bar{c}$ transforming as
\begin{align}
\bar{c}(x) \to \bar{c}(x) - \dfrac{\theta}{e\,\xi} \,(\partial_{\mu}A^{\mu})
\label{Ghost}
\end{align}
then the scalar QED Lagrangian \reff{QED} is invariant \textit{without having to use the equation of motion for $c$}. This is BRST invariance. The crucial point  is that under the BRST transformations given by \reff{Gauge} and \reff{Ghost}, the Lagrangian is invariant despite the presence of the gauge-fixing term $(\partial_{\mu}A^{\mu})^2$. 

We now turn to pure $R^2$ gravity. Consider the action 
 \begin{align}
S=\int d^4x \sqrt{-g}\,(\alpha \,R^2 +\bar{c}\, \Box c)
\label{Rsquared}
\end{align}
where again $c(x)$ and $\bar{c}(x)$ are independent Grassmann-valued fields. This action is not Weyl-invariant i.e. it is not invariant under the transformation $g_{\mu\nu} \to \Omega^2(x) g_{\mu\nu}$ where $\Omega(x)$ is an arbitrary smooth function. Consider now the Weyl transformation
\beq
g_{\mu\nu}\to e^{2 \,\theta\, c(x)} g_{\mu\nu} = (1+2 \,\theta \,c)\,g_{\mu\nu}  
\eeq{Grass}
where $\theta$ is again an arbitrary Grassmann number. Under this transformation we have 
\begin{align}
\sqrt{-g}\, \alpha \,R^2 \to \sqrt{-g} \,(\alpha\,R^2 -12 \,\alpha\,R \,\theta\, \Box c\, )
\label{Gravity}
\end{align}
where the following transformations were used: $\sqrt{-g} \to (1+ 4 \,\theta\, c)\,\sqrt{-g}$ and $ R \to (1-2 \,\theta \,c)\, R - 6 \,\theta \,\Box c$. Again, we used that $\theta^2=0$. Under the transformation \reff{Grass}, $\Box c$ transforms as
\beq
\Box c \to (1- 2 \,\theta \,c)\, \Box c
\eeq{Boxc}
where $g^{\mu\nu} \partial_{\mu} c \,\partial_{\nu} c=0$ was used (this stems from the fact that $g^{\mu\nu}$ is symmetric and $c$ is Grassmann). The equation of motion for 
$c$ is $\Box c=0$ and we see from \reff{Gravity} that $\sqrt{-g} \,\alpha\,R^2$ is Weyl invariant on-shell. However, we can dispense with the on-shell condition if we also allow $\bar{c}$ to transform as
\begin{align}
\bar{c} \to (1-2 \,\theta \,c) \,\bar{c} + 12 \,\alpha\,R \,\theta \,.
\label{Cbar}
\end{align}
We then obtain
\begin{align}
\sqrt{-g}\,\bar{c} \,\Box c \to \sqrt{-g}\,(\bar{c} \,\Box c + 12\, \alpha\,R \,\theta\,\Box c)\,.
\label{Cbar2}
\end{align}
The last term on the right hand side of \reff{Cbar2} above cancels precisely the last term on the right hand side of  \reff{Gravity}. Therefore, the action \reff{Rsquared} is invariant under the combined transformations of \reff{Grass} and \reff{Cbar} (which we refer to to as BRST transformations). \textit{ This is the BRST invariance of pure $R^2$ gravity}.  Note that BRST invariance does not require any on-shell or restricted Weyl condition. It is a generalization of Weyl (conformal) invariance that is valid in the presence of the Weyl-breaking $R^2$ term.  

Let us now take a closer look at what is common and what is different between the BRST invariance of pure $R^2$ gravity and the BRST invariance in the gauge theories of particle physics (for concreteness and simplicity, we will consider scalar QED again but the main points apply also to QCD).  The BRST invariance in scalar QED can be viewed as a generalization of  \textit{gauge invariance} in the presence of the gauge-fixing (and hence gauge-breaking) term $(\partial_{\mu}\,A^{\mu})^2$. The are two points in common between the scalar QED and  $R^2$ cases. First, the Ricci scalar $R$ under a Weyl transformation and the term $\partial_{\mu}\,A^{\mu}$ under a gauge transformation both pick up an extra $\Box \Phi(x)$ term (where $\Phi(x)$ represents either a conformal factor $\Omega(x)$ in a Weyl transformation or a function $f(x)$ in a gauge transformation). Recall that in a BRST transformation, $\Phi(x)$ is a product of a Grassmann number $\theta$ with a Grassmann field (the product yields a commuting (bosonic) quantity).  The second point in common is that $R$ and $\partial_{\mu}\,A^{\mu}$ are both \textit{squared}. The squaring yields a $(\Box \Phi(x))^2$ term which is \textit{zero} since $\theta^2=0$. The squaring still leaves one extra $\Box \Phi(x)$ term and this is cancelled out in both cases via the transformation property of a Grassmann field. These two common points render the analogy between the two cases quite strong. However, there is one important difference. In scalar QED (and in QCD) , the BRST transformations are associated with \textit{gauge transformations}. The BRST invariance of pure $R^2$ gravity that we are considering here is not associated with gauge transformations (such as diffeomorphisms) but with Weyl (local scale) transformations. We will see that this difference plays an important role when the theory is quantized.     

\section{Spontaneous breaking of BRST symmetry}

We now show that the BRST-invariant action
\begin{align}
S=\int d^4x \sqrt{-g}\,(\alpha R^2 +\bar{c}\, \Box c)
\label{R4}
\end{align}
is conformally equivalent to an action that involves the Einstein-Hilbert term; this will involve the spontaneous breaking of BRST symmetry. The starting point is to introduce a auxiliary field $\sigma(x)$ to rewrite the above action into the equivalent form 
\begin{align}
S_1&=\int d^4x \sqrt{-g}\,(-\alpha (b\,\sigma +R)^2 + \alpha R^2+ \bar{c}\, \Box c)\nonumber\\
& \int d^4x \sqrt{-g}\,(-\alpha \,b^2\,\sigma^2  -2\, \alpha\,b\, R\,\sigma+ \bar{c}\, \Box c)
\label{Requiv}
\end{align}
where $b$ is a real \textit{non-zero} constant with dimensions of mass squared and $\sigma(x)$ is dimensionless. Action \reff{Requiv} is equivalent to the original action \reff{R4} since adding the squared term in the first line of \reff{Requiv} does not alter anything (classically, the  equations of motion are unaffected and quantum mechanically, the path integral over $\sigma$  is a Gaussian which yields a constant). The equivalent action \reff{Requiv} is also BRST invariant; it is invariant under the following transformations:  
\begin{align}
g_{\mu\nu}\to (1+2\,\theta\,c)\,g_{\mu\nu}\quad;\quad \bar{c}\to  (1-2\,\theta\,c)\, \bar{c}-12\,\theta\,\alpha\,b\,\sigma\quad;\quad \sigma\to 
(1-2\theta\,c)\, \sigma
\label{Trans}
\end{align}
where $\theta$ is again a Grassmann number. Note that the BRST invariance requires the auxiliary field $\sigma$ to transform besides the fields $g_{\mu\nu}$ and  $\bar{c}$. We now perform the following conformal (Weyl) transformation:
\begin{align}
g_{\mu\nu}\to \sigma^{-1}\,g_{\mu\nu} \nonumber\\
\bar{c}\to \sigma\,\bar{c}
\label{Conf}
\end{align}
which leads to $\sqrt{-g}\to\sigma^{-2}\,\sqrt{-g}$ and $R\to \sigma\,R-6\,\sigma^{3/2}\Box (\sigma^{-1/2})$. 
Under the above conformal transformation, action \reff{Requiv} becomes
\begin{align}
S_2&=\int d^4x \sqrt{-g}\,(-\alpha\,b^2-2 \,\alpha\,b\,R+ \dfrac{3\alpha\,b}{\sigma^2}\, \partial_{\mu}\sigma \, \partial^{\mu}\sigma
+\bar{c}\, \Box c - \dfrac{1}{\sigma}\,\bar{c}\, \,\partial^{\mu}c \,\partial_{\mu}\sigma)\,.
\label{REin}
\end{align}
The above  action is no longer invariant under the BRST transformations given by \reff{Trans}. The BRST symmetry has been spontaneously broken. The factor $\sigma^{-1}$ appearing in the conformal transformation \reff{Conf} is valid only for non-zero 
$\sigma$ so that the VEV (vacuum expectation value) of the field $\sigma$ must  be non-zero. The VEV is therefore not invariant under the BRST transformation $ \sigma\to (1-2\theta\,c)\, \sigma$ leading to the spontaneous breaking of the BRST symmetry. 

We can identify $-2 \,\alpha\,b\,R$ as an Einstein-Hilbert term if we equate $-2 \,\alpha\,b $ with $ \frac{1}{16\pi\,G}$ where $G$ is Newton's constant. The constant term $-\alpha\,b^2$ can then be associated with a cosmological constant $\Lambda=-b/4$. Note that though $-2\,\alpha\,b$ is positive, the constant $b$ can be either positive or negative (but not zero). This implies that the cosmological constant can be either positive corresponding to a de Sitter (dS) background or negative corresponding to an anti-de Sitter (AdS) background but it cannot be identically zero. We can then express \reff{REin} as the following Einstein action,
\begin{align}
S_E&=\int d^4x \sqrt{-g}\,\Big(\frac{1}{16\pi\,G}(R-2\,\Lambda)+ \dfrac{3\alpha\,b}{\sigma^2}\, \partial_{\mu}\sigma \, \partial^{\mu}\sigma
+\bar{c}\, \Box c - \dfrac{1}{\sigma}\,\bar{c}\, \,\partial^{\mu}c \,\partial_{\mu}\sigma)\Big)\,.
\label{REin2}
\end{align}
We have left the constant  $3\, \alpha\,b$ in the action for simplicity but it is not an independent constant; it is equal to $\tfrac{-3}{32\,\pi G}$. We therefore obtain an Einstein-Hilbert action with non-zero cosmological constant, a kinetic term for the scalar $\sigma$ (which we will express in canonical form later) and an interaction term. Recall that $\sigma$ is non-zero so that divisions by $\sigma$ pose no issue. It is well-known that in spontaneously broken theories, the vacuum breaks the symmetry but it is not actually broken in the Lagrangian but manifested or realized in a different way \cite{Schwartz}.  It can be directly verified (see Appendix A) that the Einstein action \reff{REin2} is invariant under the following transformations:
\begin{align}
&\sigma\to (1-2\theta\,c)\, \sigma \,\,,\,\,g_{\mu\nu}\to g_{\mu\nu}\text{ and } \bar{c}\to \bar{c}-12\,\theta\,\alpha\,b\,.
\label{Trans3}
\end{align}
The BRST symmetry of action \reff{Requiv} manifests itself in the Einstein action \reff{REin2} via its symmetry under the above transformations \reff{Trans3}. We now show how transformation \reff{Trans3} stems from the  BRST  transformations \reff{Trans}. In the Einstein action and transformation  \reff{Trans3} label the metric and the barred Grassmann field with a subscript E i.e.  $g_{{\mu\nu}_E}$ and $\bar{c}_E$.  In action \reff{Requiv} and transformation \reff{Trans} we leave $g_{\mu\nu}$ and $\bar{c}$ as is. Then the conformal transformation \reff{Conf} yields $g_{{\mu\nu}_E}=\sigma\,g_{\mu\nu}$ and $\bar{c}_E= \sigma^{-1}\,\bar{c}$. Under the BRST transformations \reff{Trans} we obtain  $g_{{\mu\nu}_E}=\sigma\,g_{\mu\nu} \to (1-2\,\theta\,c)\,\sigma \,(1+2\,\theta\,c)\,g_{\mu\nu}=\sigma\,g_{\mu\nu} =g_{{\mu\nu}_E}$ and $ \bar{c}_E=\sigma^{-1}\,\bar{c} \to  (1+2\,\theta\,c)\,\sigma^{-1}\big( (1-2\,\theta\,c)\, \bar{c}-12\,\theta\,\alpha\,b\,\sigma\big)= \sigma^{-1} \bar{c}-12\,\theta\,\alpha\,b=\bar{c}_E-12\,\theta\,\alpha\,b$. We have therefore obtained the transformations $g_{{\mu\nu}_E} \to g_{{\mu\nu}_E}$ and $ \bar{c}_E\to \bar{c}_E-12\,\theta\,\alpha\,b$ which correspond to those in \reff{Trans3}. Note that we used $ \sigma\to (1-2\theta\,c)\, \sigma$  in \reff{Trans} to derive this, so the transformation of $\sigma$ is also part of \reff{Trans3}.    

We can define a real massless scalar field $\psi(x)=\sqrt{-3 \alpha\,b}\,\ln{\sigma(x)}$ so that the kinetic term for $\sigma$ is expressed in canonical form. The Einstein action \reff{REin2} expressed in terms of the field $\psi$ is
 \begin{align}
S&=\int d^4x \sqrt{-g}\,\Big(\frac{1}{16\pi\,G}(R-2\,\Lambda)- \partial_{\mu}\psi \, \partial^{\mu}\psi
+\bar{c}\, \Box c - \dfrac{1}{\sqrt{-3 \alpha\,b}}\,\bar{c}\, \partial^{\mu}c \,\partial_{\mu}\psi\Big)\,.
\label{REin3}
\end{align}
The massless scalar field $\psi$ corresponds to the Nambu-Goldstone boson of the broken sector. Under transformation  \reff{Trans3}, the field $\psi$ transforms as a shift $\psi \to \psi-\sqrt{-3 \alpha\,b}\,2\, \theta\,c$ (whereas $\bar{c}\to \bar{c}-12\,\theta\,\alpha\,b$
and $g_{\mu\nu}\to g_{\mu\nu}$). The above action \reff{REin3} is invariant under those transformations (see Appendix A). This is in accord with what we expect from spontaneously broken theories: the original symmetry in the Lagrangian manifests itself in the broken sector as a shift symmetry of the Goldstone bosons \cite{Schwartz}.  

\subsection{Quantum anomaly} 

We saw that the action \reff{Requiv} is BRST invariant under the following transformations:\\$g_{\mu\nu}\to (1+2\,\theta\,c)\,g_{\mu\nu}\,\,,\,\,\bar{c}\to  (1-2\,\theta\,c)\, \bar{c}-12\,\theta\,\alpha\,b\,\sigma \,\,,\,\,\sigma\to 
(1-2\theta\,c)\, \sigma$. Each transformation involves a Weyl transformation where the conformal factor is expressed in terms of of a product of two Grassmann variables The BRST symmetry is therefore a generalization of Weyl (conformal) symmetry. After quantization, renormalization introduces a scale which breaks the BRST symmetry since it automatically breaks Weyl symmetry (leading to the well-known Weyl (conformal) anomaly). So the BRST symmetry of pure $R^2$ gravity is anomalous.  This is in contrast to the BRST invariance of gauge theories like QCD which have no anomaly. 

After the BRST symmetry is spontaneously broken and we obtain the Einstein action \reff{REin2}, we saw that the BRST symmetry  manifests itself now in the Einstein action as a symmetry under the transformations \reff{Trans3}. This symmetry is also anomalous since the transformation of the field $\sigma$ is a Weyl transformation and renormalization breaks this symmetry (leading again to the Weyl (conformal anomaly). Another way to see this is to note that the only fields that transform in \reff{Trans3} are $\bar{c}$ and $\sigma$. The transformation for $\bar{c}$  is simply a constant shift  so that its path integral measure  $\mathcal{D} \bar{c}$ is invariant.  However, $\sigma$ undergoes a Weyl transformation and this introduces a non-trivial Jacobian J (i.e. $J\ne 1$) to the measure  $\mathcal{D} \sigma$. Since the measure is not invariant, this implies there is an anomaly \cite{Fujikawa}.  So the symmetry in the unbroken phase and its associated symmetry in the broken phase are both anomalous. Our finding here is in accord with previous work that shows that when the Weyl or conformal symmetry is spontaneously broken there is conformal anomaly matching between the unbroken and broken phases \cite{Schwimmer,Bizet}.

\section{Conclusion} 

In the last six years or so, we have kept discovering new aspects of pure $R^2$ gravity. A non-exhaustive list includes its unitarity among quadratic gravity theories \cite{Lust3}, its conformal equivalence to Einstein gravity with non-zero cosmological constant and massless scalar field \cite{Lust1,Lust2,AE1,Lust3,AE2}, its restricted Weyl symmetry \cite{AE3,Oda1,Oda2}, its spontaneous symmetry breaking to Einstein gravity \cite{AE1,AE2} and the lack of a propagating graviton when the theory is linearized about a Minkowski background \cite{Lust3} (where there is no Einstein equivalence since the cosmological constant is zero).  

In this paper, we have gained further insights into this theory. We saw that pure $R^2$  gravity has an analog with the gauge-fixing term  $(\partial_{\mu}A^{\mu})^2$ in gauge theory. $R^2$ is not invariant under the Weyl transformation $g_{\mu\nu} \to \Omega^2 (x)\,g_{\mu\nu}$ just like $(\partial_{\mu}\,A^{\mu})^2$ is not invariant under the gauge transformation $A_{\mu}\to A_{\mu} + \frac{1}{e} \,\partial_{\mu} f(x)$. However, each have a residual symmetry (when $\Box \Omega=0$ is satisfied in the gravity case and $\Box f=0$ is satisfied in the gauge case). This analogy opened the door towards enlarging the symmetry of pure $R^2$ gravity to include BRST symmetry.  We first showed that when a massless scalars field was included in the pure $R^2$ action, the condition $\Box \Omega=0$ could be met automatically when the equations of motion were satisfied i.e. we went from restricted Weyl to on-shell Weyl invariance. Finally, we obtained the BRST invariance of pure $R^2$ gravity where no restricted Weyl or on-shell condition is required. The BRST transformations involve Weyl transformations where the conformal factor is composed of products of Grassmann variables (the conformal factor itself is commutative).  The important point is that the BRST invariance exists despite the Weyl-breaking $R^2$ term. 

There is one important difference between the BRST symmetry in gauge theories like QCD and the BRST symmetry that we have considered here for pure $R^2$ gravity. Gauge invariance in particle physics is preserved after quantization. The BRST invariance of QCD is a generalization of gauge invariance so that it is also preserved after quantization; there is no anomaly. In contrast to gauge symmetry, global scale or Weyl (local scale) symmetry is broken after quantization since renormalization introduces a scale. The BRST symmetry of pure $R^2$ gravity is a generalization of Weyl (conformal) symmetry so that it is also broken after quantization leading to the well-known Weyl (conformal) anomaly. After the spontaneous breaking of the BRST symnmetry, we obtained an Einstein action. We showed that this action has its own symmetry and that it is also anomalous. This is in accord with previous work that shows that when the Weyl (conformal) symmetry is spontaneously broken there is conformal anomaly matching between the unbroken and broken sectors \cite{Schwimmer, Bizet}. 

The focus of this paper was pure $R^2$ gravity because of its many special and attractive features that we previously mentioned. All other quadratic gravity theories (like Weyl-squared, Riemann-squared, etc.), apart from boundary terms,  can be expressed as a linear combination of $R^2$ and $R^{\mu\nu}R_{\mu\nu}$.  The latter term, the square of the Ricci tensor, appears in quantum corrections to General Relativity (GR) and even though it does not constitute a valid UV completion of GR due to its non-unitarity (yields a massive spin two ghost \cite{Lust3,Stelle}), it still makes a well-known calculable short-range correction to the Newtonian potential \cite{Donoghue, Schwartz}.  Like $R^2$, the term $R_{\mu\nu}R^{\mu\nu}$ is not Weyl-invariant so it would be of interest to see if it can be BRST invariant. It is not in the form of a scalar squared like  $(\partial_{\mu}A^{\mu})^2$ or $R^2$, so one may be inclined to think that the BRST formalism would not apply here.  However, like pure $R^2$, it was shown in \cite{AE3} that  $R_{\mu\nu}R^{\mu\nu}$ is restricted Weyl invariant  (up to a boundary term). This suggests that the procedure used to establish the  BRST invariance of pure $R^2$ gravity might in the end also work for this quadratic theory. It would therefore be worthwhile and interesting to investigate this further.

\section*{Acknowledgments}
A.E. acknowledges support from a discovery grant of the National Science and Engineering Research Council of Canada (NSERC).

\begin{appendices}
\numberwithin{equation}{section}
\setcounter{equation}{0}

\section{Symmetry of Einsten Action }
\numberwithin{equation}{section}
\setcounter{equation}{0}
In this appendix we show that the Einstein action \reff{REin2} given by
\begin{align}
S_E&=\int d^4x \sqrt{-g}\,\Big(\frac{1}{16\pi\,G}(R-2\,\Lambda)+ \dfrac{3\alpha\,b}{\sigma^2}\, \partial_{\mu}\sigma \, \partial^{\mu}\sigma
+\bar{c}\, \Box c - \dfrac{1}{\sigma}\,\bar{c}\, \,\partial^{\mu}c \,\partial_{\mu}\sigma)\Big)
\label{REin5}
\end{align}
is invariant under the transformations \reff{Trans3} given by
\begin{align}
\sigma\to (1-2\theta\,c)\, \sigma \,\,,\,\,g_{\mu\nu}\to g_{\mu\nu}\text{ and } \bar{c}\to \bar{c}-12\,\theta\,\alpha\,b\,.
\label{Trans5}
\end{align}
Under the above transformation, the metric $g_{\mu\nu}$ does not change so that $\sqrt{-g}$ as well as the term $ \sqrt{-g}\,\frac{1}{16\pi\,G}(R-2\,\Lambda)$ does not change. The other terms in the above Einstein action transform as
\begin{align}
&\dfrac{3\alpha\,b}{\sigma^2}\, \partial_{\mu}\sigma \, \partial^{\mu}\sigma \to\dfrac{3\alpha\,b}{\sigma^2}\, \partial_{\mu}\sigma \, \partial^{\mu}\sigma-\frac{12\,\theta\,\alpha\,b}{\sigma}\, \partial^{\mu}c\,\partial_{\mu}\sigma\nonumber\\
&- \dfrac{1}{\sigma}\,\bar{c}\, \,\partial^{\mu}c \,\partial_{\mu}\sigma)\to- \dfrac{1}{\sigma}\,\bar{c}\, \,\partial^{\mu}c \,\partial_{\mu}\sigma +\frac{12\,\theta\,\alpha\,b}{\sigma}\, \partial^{\mu}c\,\partial_{\mu}\sigma\nonumber\\
&\bar{c}\, \Box c \to \bar{c}\, \Box c -12\, \theta\,\alpha\,b \,\,\Box c
\label{Trans6}
\end{align}
where we used that $\theta^2=0$ (since $\theta$ is a Grassmann number) and that $g^{\mu\nu} \,\partial_{\mu}c\,\partial_{\nu} c=0$ since $g^{\mu\nu}$ is symmetric and $c(x)$ and its derivatives are Grassmann fields. We see that the extra term $ -\frac{12\,\theta\,\alpha\,b}{\sigma}\, \partial^{\mu}c\,\partial_{\mu}\sigma$ in the first line of \reff{Trans6} is canceled exactly by the extra term in the second line. The extra term in the third line, $-12\, \theta\,\alpha\,b \,\,\Box c$, where $-12\, \theta\,\alpha\,b$ is a constant, does not cancel out with any other extra term in \reff{Trans6}.   However,  $\sqrt{-g}\,\Box c$ is a total derivative that yields an inconsequential boundary term in the action. We have therefore shown that action \reff{REin5} is invariant under transformations \reff{Trans5}.

We saw in section 4 that the Einstein action \reff{REin5} can be expressed in terms of a real massless scalar field $\psi(x)=\sqrt{-3 \alpha\,b}\,\ln{\sigma(x)}$ as action \reff{REin3}:
 \begin{align}
S&=\int d^4x \sqrt{-g}\,\Big(\frac{1}{16\pi\,G}(R-2\,\Lambda)- \partial_{\mu}\psi \, \partial^{\mu}\psi
+\bar{c}\, \Box c - \dfrac{1}{\sqrt{-3 \alpha\,b}}\,\bar{c}\, \partial^{\mu}c \,\partial_{\mu}\psi\Big)
\label{REin6}
\end{align}
where $\psi$ was identified as the Nambu-Goldstone boson of the broken sector. We stated in section 4 that the action \reff{REin6} was invariant under the following transformations:
\begin{align}
\psi \to \psi-\sqrt{-3 \alpha\,b}\,2\, \theta\,c \text{ , } \bar{c}\to \bar{c}-12\,\theta\,\alpha\,b \text{ and } g_{\mu\nu}\to g_{\mu\nu}\,.
\label{Psi2}
\end{align}
We now verify this statement. Under \reff{Psi2} the last three terms in action \reff{REin6} transform as:
 \begin{align}
&- \partial_{\mu}\psi \, \partial^{\mu}\psi \to - \partial_{\mu}\psi \, \partial^{\mu}\psi + 4 \,\theta \sqrt{-3\,\alpha\,b} \,\,\partial_{\mu}\psi\,  \partial^{\mu}c\nonumber\\
&- \dfrac{1}{\sqrt{-3 \alpha\,b}}\,\bar{c}\, \partial^{\mu}c \,\partial_{\mu}\psi \to - \dfrac{1}{\sqrt{-3 \alpha\,b}}\,\bar{c}\, \partial^{\mu}c \,\partial_{\mu}\psi - 4 \,\theta \,\sqrt{-3\,\alpha\,b}\,\, \partial_{\mu}\psi\,  \partial^{\mu}c\,\nonumber\\
&\bar{c}\, \Box c \to \bar{c}\, \Box c -12 \,\theta\,\alpha\,b\,\Box c\,.
\label{Trans4}
\end{align}
We see that the extra term $+ 4 \,\theta\, \sqrt{-3\,\alpha\,b} \,\,\partial_{\mu}\psi\,  \partial^{\mu}c$ in the first line above is cancelled by the extra term on the second line which is equal to its negative. The only extra term that is not cancelled is the term  $-12 \,\theta\,\alpha\,b\,\Box c$ appearing in the last line. However, $\sqrt{-g}\,\, \Box c$ is a total derivative which yields a boundary term with no physical consequence. We have therefore verified that the Einstein action \reff{REin6} is indeed invariant under the transformations \reff{Psi2}. 
\end{appendices}

\end{document}